\documentstyle[preprint,tighten,eqsecnum,floats,epsfig,aps]{revtex}

\begin{document}
\draft

\title{Active and Passive Quantum Erasers for Neutral Kaons}

\author{A. Bramon$^{1}$, G. Garbarino$^{2}$ and B.~C.~Hiesmayr$^{3}$}

\address{$^1$Grup de F{\'\i}sica Te\`orica,
Universitat Aut\`onoma de Barcelona, E--08193 Bellaterra, Spain}

\address{$^{2}$Dipartimento di Fisica Teorica, Universit\`a di Torino
and INFN, Sezione di Torino,\\ I--10125 Torino, Italy}

\address{$^{3}$Institute for Experimental Physics, University of Vienna,
A--1090 Vienna, Austria}
\date{\today}
\maketitle

\begin{abstract}
Quantum marking and quantum erasure are discussed for the neutral kaon system.
Contrary to other two--level systems, strangeness and lifetime of a neutral kaon
state can be alternatively measured via an ``\emph{active}" or a ``\emph{passive}"
procedure. This offers new quantum erasure possibilities. In particular,
the operation of a quantum eraser in the ``\emph{delayed choice}" mode is clearly
illustrated.
\end{abstract}


\newpage
\pagestyle{plain}
\baselineskip 16pt
\vskip 48pt

\section{Introduction}

The introduction of \emph{quantum marking} and \emph{erasure}  by
Scully and Dr\"uhl \cite{scully82,scully91} in 1982 opened the
possibility  to discuss some of the most relevant and subtle
aspects of quantum measurement. Indeed, since the foundation of
quantum mechanics, one is well aware of the crucial role played by
measurement devices; a role which, occasionally, is even extended
to conscious observers, as in the well known proposal by Wigner.
Quantum marking and erasure are useful tools to investigate these
basic issues. Up to now, experimental tests of quantum erasure
have been performed with atom interferometers \cite{Durr} and entangled photon pairs
produced via spontaneous parametric down--conversion (SPDC)
\cite{zeilinger95,scully00,Ts00,walborn,Tr02,Kim03}. 
The general purpose of the present paper is to extend these quantum eraser
considerations to entangled massive particles, i.e., $K^0\bar K^0$
pairs produced in $\phi$--resonance decays \cite{handbook} or
proton--antiproton annihilations at rest \cite{CPLEAR}.

The basic idea behind quantum marking and erasure is that two
indistinguishable ---and thus potentially interfering---
amplitudes of a quantum system, the \emph{object}, can be made
distinguishable thanks to the \emph{entanglement} between the
object and a second quantum system, the \emph{meter}, which thus
carries a kind of quantum \emph{mark}. In ordinary two--path
interferometric devices, the latter, marked system is frequently
called  a ``which way'' detector. If the information stored in the
``which way'' detector is even in principle accessible, the object
system looses all its previous interference abilities. However, if
one somehow manages to ``erase'' the meter mark and thus the
distinguishability of the object amplitudes, the original object
interference effects can be restored. This is achieved by
correlating the outcomes of the measurements on the object system
with those of suitable erasing measurements on the meter system.

The case in which the meter is a system distinct and spatially
separated from the object is of particular interest. Indeed, in
this case the decision to erase or not the meter mark and the
distinguishability of the object amplitudes ---and therefore to
observe or not interference--- can be taken long after the
measurement on the object system has been completed. Quantum
erasure is then performed in the so called \emph{delayed choice}
mode which best captures the essence and most subtle aspects of
this phenomenon \cite{scully82,scully91,mohrhoff,scully99}. This
mode of quantum erasure can be applied to entangled $K^0\bar K^0$
pairs as well as to SPDC photon pairs, but, specific of the kaon
case is the possibility that the erasure operation can be carried
out via two distinct procedures: either \emph{actively}, i.e., by
exerting the free will of the experimenter, or \emph{passively},
i.e., randomly exploiting a particular quantum--mechanical
property of the meter system. This opens new possibilities for
quantum erasure with kaons.

The first specific purpose of this paper is to discuss the
existence of these two different active and passive
quantum erasure procedures for neutral kaons. A second, twofold purpose is
to show both the simplicity of the \emph{delayed choice} mode of
quantum erasure and its extension to \emph{passive} measurements
when operated with neutral kaons. These kaons turn
out to be most suitable for achieving these two purposes.

\section{Measurements on neutral kaons}
\label{measurements}

Contrary to what happens with other two--level quantum systems such as
spin--$1/2$ particles or photons, neutral kaons only
exhibit two different measurement bases \cite{bg1,ABN}: the strangeness and the
lifetime bases.

The strangeness basis, $\lbrace K^0,\bar K^0\rbrace$ with $\langle K^0|\bar
K^0\rangle=0$, is the appropriate one to discuss strong production and reactions of
kaons. If a dense piece of nucleonic matter is inserted along a
neutral kaon beam, the incoming state is projected either into a $K^0$ by the
strangeness conserving strong interaction $K^0 p\rightarrow K^+ n$ or into
a $\bar K^0$ via $\bar K^0 p\rightarrow \Lambda \pi^+$, $\bar K^0 n\rightarrow \Lambda \pi^0$ 
or $\bar K^0 n\rightarrow K^-p$. These strangeness detections are
totally analogous to the projective von Neumann measurements of two--channel
analyzers for polarized photons or Stern--Gerlach setups for spin--$1/2$ particles. By
inserting the piece of matter along a kaon beam, one induces an \emph{active}
measurement of strangeness.

The strangeness content of neutral kaon states
can be alternatively determined by observing their semileptonic decay modes.
Indeed, these semileptonic decays obey the well tested $\Delta S=\Delta
Q$ rule which allows the modes
\begin{equation}
K^0(\bar s d) \to \pi^-(\bar u d)\;+\;l^+\;+\;\nu_l \; , \; \;
\bar{K}^0(s\bar d) \to\pi^+(u \bar d)\;+\;l^-\;+\;\bar\nu_l ,
\end{equation}
where $l$ stands for $e$ or $\mu$, but forbids decays into the
respective charge conjugated modes. Obviously, the experimenter
cannot induce a kaon to decay semileptonically and not even at a given
time: he or she can only sort at the end of the day all observed
events in proper decay modes and time intervals.
We call this discrimination between $K^0$ and $\bar K^0$ via the
identification of the kaon semileptonic decay modes
a  \emph{passive} measurement of strangeness.

The detection efficiency for both \emph{active} and \emph{passive} strangeness measurements
is rather limited. In the former case, the detection efficiency is close to $1$ only
for ultrarelativistic kaons; indeed, by Lorentz contraction the piece of matter
is seen by the incoming kaon as extremely dense and kaon--nucleon strong interactions become
much more likely than kaon weak decays. In the case of \emph{passive} strangeness
detection, the efficiency is given by the $K_L$ and $K_S$ semileptonic branching
ratios, which are $\simeq 0.66$ and $\simeq 1.1\times 10^{-3}$, respectively. These limited
detection efficiencies, which originate a serious problem \cite{bg1,gen2004}
when discussing Bell--type
tests for entangled kaons \cite{ABN,GGW,eberhard,kaon-bell,BH1}, do not introduce any
conceptual difficulty here. They simply represent a practical difficulty
in obtaining large statistical samples.

The second basis, the lifetime basis $\{K_S,K_L\}$, consists of
the short-- and long--lived states having well defined masses
$m_{S(L)}$ and decay widths $\Gamma_{(S)L}$. It is the appropriate
basis to discuss free space  propagation, with:
\begin{eqnarray}\label{timeevolution}
|K_{S}(\tau)\rangle = e^{-i\lambda_{S}\tau} |K_{S}\rangle ,\;\;
|K_{L}(\tau)\rangle = e^{-i\lambda_{L}\tau} |K_{L}\rangle ,
\end{eqnarray}
and $\lambda_{S(L)}=m_{S(L)}-i\Gamma_{S(L)}/2$. These states
preserve their own identity in time, but, since $\Gamma_S\simeq
579 \, \Gamma_L$ \cite{PDG}, the $K_S$ component of a neutral kaon
extincts much faster than the $K_L$ component. To observe if a
kaon is propagating as a $K_S$ or $K_L$ at (proper) time $\tau$,
one has to identify at which time it subsequently decays. 
Kaons which show a prompt decay, i.e., a decay
between $\tau$ and $\tau + \Delta \tau$, have to be identified as $K_S$'s, while those
decaying later than $\tau + \Delta \tau$ have to be identified as $K_L$'s. The
probabilities for wrong $K_S$ or $K_L$ identification are then given by
$\exp(- \Gamma_S\, \Delta \tau)$ and $1 - \exp(- \Gamma_L\, \Delta \tau)$, respectively.
With $\Delta \tau = 4.8\, \tau_S$, both $K_S$ and $K_L$ misidentification probabilities
are equal and reduce to $\simeq 0.8$\% \cite{bg1,eberhard,BGH1}.
Such a procedure, in which the experimenter chooses to allow for
free space propagation, represents  an \emph{active}
measurement of lifetime.

Since four decades one knows that the neutral kaon system violates
$CP$ symmetry. Among other things, this implies that the weak
interaction eigenstates are not strictly orthogonal to each other, 
$\langle K_S|K_L\rangle= 2 ({\rm
Re}\,\varepsilon)/(1+|\varepsilon|^2)\simeq 3.2 \cdot10^{-3}$
\cite{PDG,kabir}. However, by neglecting these small $CP$
violation effects one can discriminate between $K_S$'s and $K_L$'s
by leaving the kaons to propagate in  free space and observing
their distinctive nonleptonic $K_S \to 2\pi$ or $K_L \to 3\pi$
decay modes.  This represents a \emph{passive} measurement of
lifetime, since the type of kaon decay modes ---nonleptonic in the
present case, instead of semileptonic as before--- cannot be in
any way influenced by the experimenter. The measurement procedure
is determined by the quantum dynamics of kaon decays.

Summarizing, we have two conceptually different experimental
procedures ---\emph{active} and \emph{passive}--- to measure each one
of the only two neutral kaon observables: strangeness or lifetime.
The \emph{active} measurement of strangeness is monitored by
strangeness conservation while the corresponding \emph{passive}
measurement is assured by the $\Delta S=\Delta Q$ rule.
\emph{Active} and \emph{passive} lifetime measurements are
efficient thanks to the smallness of $\Gamma_L/\Gamma_S$ and
$\varepsilon$, respectively.
The existence of these two procedures is of no interest when testing Bell
inequalities with kaons, where \emph{active} measurements must be
considered \cite{bg1,ABN}. 
However, it opens new possibilities for kaonic quantum
erasure experiments which have no analog for any other two--level
quantum system considered up to date.

\section{Single kaons: time evolution and measurements}
\label{evolution}

Let us start discussing the time evolution of a
single neutral kaon which is initially produced (say) as a
$K^0$. Just after production, at $\tau=0$, it is described by
the equal superposition of the two lifetime eigenstates,
$|K^0(0)\rangle=\lbrace |K_S\rangle+|K_L\rangle\rbrace/\sqrt{2}$,
and, according to Eq.~(\ref{timeevolution}), it starts propagating
in free space in the coherent superposition:
\begin{eqnarray}
\label{singletau}
|K^0(\tau)\rangle=\frac{1}{\sqrt{2}}\lbrace
e^{-i\lambda_S \tau}|K_S\rangle+e^{-i\lambda_L \tau}|K_L\rangle\rbrace\;.
\end{eqnarray}

Note that the state $|K^0(\tau)\rangle$ evolves through a single
spatial trajectory comprising
\emph{automatically} ---i.e., with no need of any kind of double--slit
apparatus---
the two differently propagating components $K_S$ and $K_L$. This generates
the well known strangeness or
$K^0$--$\bar K^0$ oscillations. Note also that, since $\Gamma_S\simeq 579\,
\Gamma_L$,
these two components are \emph{intrinsically} marked by their remarkably
distinct decay widths.
At $\tau=0$ this lifetime mark is inoperative and there is no information on
which
component is actually propagating; $K_S$--$K_L$ interference and
thus $K^0$--$\bar K^0$ oscillations are maximal around $\tau=0$.
However, for kaons surviving after some time $\tau$,  $K_S$
propagation is known to be less likely than $K_L$ and partial
``which width'' information is obtained; interference and oscillation
phenomena are thus diminished.
This happens in the  same way as when one
obtains ``which way'' information in two--path
interferometric devices, as recently discussed
in Ref.~\cite{BGH3} in terms of quantitative
complementarity.
This dedicated analysis shows that a
variety of double--slit experiments with photons
or neutrons, as well as Mott scattering experiments with
identical nuclei, admit the very same description
as the present oscillations of the neutral kaon system.
But two advantages of the latter system are obvious:
it does not need any kind of double--slit device nor
any marking procedure; indeed,
both features are inherently supplied by Nature.

\emph{Active} and \emph{passive}  measurements of strangeness or
lifetime on single kaon states evolving, for instance, as in
Eq.~(\ref{singletau}) offer no difficulties and have been performed
since long time ago.

For the two possible \emph{active} measurements
one either inserts or not a dense piece of matter
at a (time--of--flight) distance $\tau$ along the kaon trajectory.
The corresponding quantum--mechanical probabilities are simply
derived by projecting Eq.~(\ref{singletau}) either into the
strangeness or into the lifetime basis. By properly
normalizing Eq.~(\ref{singletau}) to undecayed kaons, the
following oscillating probabilities for $K^0$ and
$\bar{K}^0$ detection at time $\tau$ are obtained:
\begin{eqnarray}
\label{k0}
&&P\left[ K^0 (\tau )\right]
= {1 \over 2} \left[ 1+{\cal V}_0(\tau )\cos (\Delta m \, \tau)\right], \\
\label{k0bar}
&&P\left[\bar{K}^0 (\tau )\right]
= {1 \over 2} \left[ 1-{\cal V}_0(\tau )\cos (\Delta m \, \tau)\right] .
\end{eqnarray}
In these expressions:
\begin{equation}
{\cal V}_0(\tau ) = \frac{1}{\cosh (\Delta \Gamma \tau/2)}
\end{equation}
is the time--dependent visibility of the strangeness oscillations,
$\Delta m \equiv m_L-m_S$ and $\Delta \Gamma \equiv \Gamma_L-\Gamma_S$.
Instead, no oscillations are predicted for lifetime measurements:
\begin{eqnarray}
\label{kl}
&&P\left[ K_L (\tau )\right]
= {1 \over 1 + e^{+ \Delta \Gamma \tau}}\; , \\
\label{ks}
&&P\left[ K_S (\tau )\right]
= {1 \over 1 + e^{-\Delta\Gamma \tau}}\; .
\end{eqnarray}

\emph{Passive} measurements of strangeness or lifetime on single
kaon states are slightly less direct. One has to measure the
$\tau$--dependent decay rate $\Gamma(f,\tau)$ which is defined as the number of
decays into the mode $f$ occurring between $\tau$ and $\tau + d\tau$
divided by $d\tau$ and by the total number of initial $K^0$'s.
Its formal expression is given by:
\begin{eqnarray}
\Gamma(f,\tau)\equiv \int d\Omega_f \, p(f,\tau) ,
\end{eqnarray}
where the integration is over the phase space of the decay
product state $f$ ($f=\pi \pi$, $\pi \pi \pi$,
$\pi^- l^+ \nu_l$ or $\pi^+ l^- \bar{\nu}_l$) and:
\begin{equation}
p(f,\tau)\equiv \frac{1}{2}\biggl|e^{-i\lambda_S\, \tau} \langle f|T|K_S\rangle
+e^{-i \lambda_L\, \tau} \langle f|T|K_L\rangle\biggl|^2 .
\end{equation}
The decay rate $\Gamma(f,\tau)$ is normalized such that
$\sum_{f}\int d\tau \, \Gamma(f,\tau)=1$
and allows one to determine the single kaon detection probabilities
through the relation:
\begin{equation}
\label{obs1sing}
P\left[K_f(\tau)\right]=\frac{\Gamma(f,\tau)}{\Gamma(K_f\to f)\, N(\tau)},
\end{equation}
which reproduces the predictions (\ref{k0})--(\ref{ks}) of \emph{active} measurements.
In Eq.~(\ref{obs1sing}), $K_f=K_S$, $K_L$, $K^0$ or $\bar{K}^0$ and
$N(\tau)\equiv [e^{-\Gamma_L\, \tau}+e^{-\Gamma_L\, \tau}]/2$
is a normalization factor taking into account the extinction with time of the beam.
Moreover, the decay widths $\Gamma(K_f\to f)$, where $K_f\to f$ stands for
the four identifying decay modes, $K_S \to \pi\pi$, $K_L \to \pi\pi\pi$,
$K^0 \to \pi^- l^+ \nu_l$ and $\bar{K}^0 \to \pi^+ l^- \bar{\nu}_l$,
are given by:
\begin{equation}
\label{singledecaywidths}
\Gamma(K_f\to f)\equiv \int d\Omega_f|\langle f|T|K_f\rangle|^2=
\frac{BR(K_L\to f)\, \Gamma_L}{|\langle
K_f|K_L\rangle|^2}= \frac{BR(K_S\to f)\, \Gamma_S}{|\langle
K_f|K_S\rangle|^2} , \nonumber
\end{equation}
where the second (third) equality is only defined for
nonvanishing values of $\langle K_f|K_L\rangle$ ($\langle
K_f|K_S\rangle$) and $BR(K_{L(S)}\to f)$ is the branching ratio for the
decay $K_{L(S)}\to f$.

The experimental equivalence of both measurement procedures and
the agreement with quantum--mechanical predictions have already been
proved \cite{PDG,CPLEARreview,BGH2}.

\section{Entangled kaon pairs: time evolution and measurements}
\label{probabilities}

Let us now consider two--kaon entangled states which are analogous
to the standard and widely used two--photon entangled states
produced by SPDC. From both $\phi$--meson resonance decays
\cite{handbook} or $S$--wave proton--antiproton annihilation
\cite{CPLEAR}, one starts at time $\tau=0$ with the state:
\begin{equation}
\label{entangled} |\phi(0)\rangle  =  \frac{1}{\sqrt 2}\left[
|K^0\rangle_l |\bar{K}^0\rangle_r - |\bar{K}^0\rangle_l
|K^0\rangle_r\right]
 =  \frac{1}{\sqrt 2}\left[
|K_L\rangle_l |K_S\rangle_r - |K_S\rangle_l |K_L\rangle_r\right] ,
\end{equation}
where $l$ and $r$ denote the ``left'' and ``right'' directions of
motion of the two separating kaons and $CP$--violating effects are neglected
in the last equality. Note that this state is
antisymmetric and maximally entangled in the two observable bases.

After production, the left  and right moving kaons evolve
according to Eq.~(\ref{timeevolution}) up to times $\tau_l$ and
$\tau_r$, respectively, thus leading to the state:
\begin{equation}
\label{notnorm}
|\phi(\tau_l,\tau_r)\rangle = \frac{1}{\sqrt
2}\left\{ e^{-i(\lambda_L \tau_l+\lambda_S
\tau_r)}|K_L\rangle_l|K_S\rangle_r -e^{-i(\lambda_S \tau_l+
\lambda_L \tau_r)}|K_S\rangle_l|K_L\rangle_r\right\} .
\end{equation}
By normalizing to kaon pairs with both members surviving up to
$(\tau_l,\tau_r)$, one obtains the state:
\begin{equation}
\label{time}
|\phi(\Delta\tau)\rangle= \frac{1}{\sqrt
{1+e^{\Delta\Gamma \Delta\tau}}}\biggl\lbrace
|K_L\rangle_l|K_S\rangle_r -e^{i \Delta m \Delta\tau} e^{{1 \over
2} \Delta \Gamma \Delta\tau}
|K_S\rangle_l|K_L\rangle_r\biggr\rbrace ,
\end{equation}
where $\Delta\tau\equiv \tau_l-\tau_r$, or, equivalently:
\begin{eqnarray}
\label{timestrangeness} |\phi(\Delta\tau)\rangle &=& \frac{1}{2
\sqrt {1+e^{\Delta\Gamma \Delta\tau}}} \left\{(1-e^{i \Delta m
\Delta\tau} e^{{1 \over 2} \Delta \Gamma \Delta\tau})
\left[|K^0\rangle_l|K^0\rangle_r-|\bar K^0\rangle_l|\bar
K^0\rangle_r\right] \right.
\nonumber \\
&& \left. +(1+e^{i \Delta m \Delta\tau} e^{{1 \over 2} \Delta
\Gamma \Delta\tau}) \left[|K^0\rangle_l|\bar K^0\rangle_r-|\bar
K^0\rangle_l|K^0\rangle_r\right] \right\} ,
\end{eqnarray}
in the strangeness basis.

Thanks to this normalization, one works with bipartite two--level
systems as for photon or spin--$1/2$ entangled pairs. For a
detailed description of the time evolution of entangled neutral
kaon pairs, see Refs.~\cite{GGW,BH1}. The analogy between state
(\ref{time}) and the polarization--entangled two--photon [idler
($i$) plus signal ($s$)] state $|\Psi \rangle = \left[ |V\rangle_i
|H\rangle_s - e^{i \Delta \phi}|H\rangle_i |V\rangle_s
\right]/\sqrt{2}$, where $\Delta \phi$ is a relative phase under
control by the experimenter, is obvious.

\subsection{\textit{Active} measurements on both kaons}

\emph{Active} joint measurements on two--kaon states are again quite
obvious. \emph{Active} strangeness measurements on both sides require strangeness
detectors inserted at $\tau_l$ and $\tau_r$. This corresponds to act with the projectors
$P_i^l P_j^r$, where $P_{i(j)}=|K^0\rangle\langle K^0|$ or $|\bar
K^0\rangle\langle \bar K^0|$, on the state
(\ref{timestrangeness}). The probabilities to observe
like-- and unlike--strangeness joint events are:
\begin{eqnarray}
\label{lSprob}
P\left[K^0(\tau_l),K^0(\tau_r)\right]=P\left[\bar{K}^0(\tau_l),\bar{K}^0(\tau_r)\right]
= \frac{1}{4}\left[1-{\cal V}(\Delta\tau) \cos(\Delta m\, \Delta \tau)\right] , && \\
\label{uSprob}
P\left[K^0(\tau_l),\bar{K}^0(\tau_r)\right]=P\left[\bar{K}^0(\tau_l),K^0(\tau_r)\right]
= \frac{1}{4}\left[1+{\cal V}(\Delta\tau) \cos(\Delta m\, \Delta
\tau)\right] , &&
\end{eqnarray}
respectively, where:
\begin{equation}
\label{visibility}
{\cal V}(\Delta\tau)=\frac{1}{\cosh(\Delta\Gamma\Delta\tau/2)}.
\end{equation}
First, we note that for $\Delta\tau=0$ we have perfect
EPR--correlations: since ${\cal V}(0)=1$, the like--strangeness
probabilities vanish and the unlike--strangeness probabilities
take the maximum value. Second, $\Delta m\, \Delta\tau$ plays the
same role as the relative orientation of polarization analyzers
for entangled photon pairs. The kaon mass difference $\Delta m$
introduces automatically a time dependent relative phase between
the two kaon amplitudes of Eq.~(\ref{time}). However, opposite to
the photon case, the visibility of Eq.~(\ref{visibility})
decreases as $|\Delta\tau|$ increases.

If one wants to measure, \emph{actively}, strangeness on the left and lifetime
on the right, one has to remove the piece of matter on the right to
allow for free kaon propagation in space.
One then measures the following non--oscillating joint probabilities:
\begin{eqnarray}
\label{probS}
&&P\left[K^0(\tau_l),K_S(\tau_r)\right]=P\left[\bar{K}^0(\tau_l),K_S(\tau_r)\right]
=\frac{1}{2\left(1+e^{\Delta\Gamma\Delta\tau}\right)} , \\
\label{probL}
&&P\left[K^0(\tau_l),K_L(\tau_r)\right]=P\left[\bar{K}^0(\tau_l),K_L(\tau_r)\right]
=\frac{1}{2\left(1+e^{-\Delta\Gamma\Delta\tau}\right)} ,
\end{eqnarray}
which are directly obtained from state (\ref{time}) by acting with the
projectors $P_i^l P_j^r$, where $P_{i(j)}=|K_S\rangle\langle K_S|$,
$|K_L\rangle\langle K_L|$, $|K^0\rangle\langle K^0|$ or
$|\bar K^0\rangle\langle \bar K^0|$.

\subsection{\textit{Passive} measurements on both kaons}

\emph{Passive} joint measurements are somewhat more involved. In
this case, one allows the entangled kaon pairs to propagate freely
in space and identifies the kaon decay \emph{times} and
\emph{modes}. As discussed in detail in Ref.~\cite{ABN}, one has
to measure the ($\tau_l$-- and $\tau_r$--dependent) joint decay rate:
\begin{eqnarray}
\Gamma(f_l,\tau_l; f_r,\tau_r)\equiv \int d\Omega_{f_l}\,
d\Omega_{f_r}\, p(f_l,\tau_l;f_r,\tau_r) ,
\end{eqnarray}
which is the obvious generalization of the previously defined
single kaon decay rate $\Gamma(f,\tau)$, with:
\begin{eqnarray}
p(f_l,\tau_l;f_r,\tau_r)&\equiv &\frac{1}{2}
\biggl|e^{-i(\lambda_L \tau_l+\lambda_S \tau_r)}
\langle f_l|T|K_L\rangle_l \langle f_r|T|K_S\rangle_r \\
&&-e^{-i(\lambda_S \tau_l+\lambda_L \tau_r)}
\langle f_l|T|K_S\rangle_l \langle
f_r|T|K_L\rangle_r\biggl|^2 . \nonumber
\end{eqnarray}
The joint decay rate $\Gamma(f_l,\tau_l;f_r,\tau_r)$ is normalized such that
$\sum_{f_l,f_r}\int d\tau_l\, d\tau_r \, \Gamma(f_l,\tau_l;
f_r,\tau_r)=1$ and supplies the joint detection
probabilities through the relation \cite{ABN}:
\begin{equation}
\label{obs1}
P\left[K_{f_l}(\tau_l),K_{f_r}(\tau_r)\right]=\frac{\Gamma(f_l,\tau_l;
f_r,\tau_r)}{\Gamma(K_{f_l}\to f_l)\, \Gamma(K_{f_r}\to f_r)\, N(\tau_l,\tau_r)} ,
\end{equation}
where $N(\tau_l, \tau_r) \equiv e^{-(\Gamma_L +\Gamma_S)(\tau_l
+\tau_r)/2} \cosh\left[\Delta\Gamma(\tau_l -\tau_r)/2\right]$ is a
normalization factor depending on both $\tau_l$ and $\tau_r$ which
accounts for the extinction of the beams, while $K_{f_l}$,
$K_{f_r}=K_S$, $K_L$, $K^0$ or $\bar{K}^0$.

In the good approximation of $CP$ conservation and the validity of the
$\Delta S=\Delta Q$ rule, it is then easy to see that the physical meaning and the
quantum--mechanical expressions for the probabilities obtained through Eq.~(\ref{obs1})
coincide with the results (\ref{lSprob})--(\ref{probL}) corresponding to
\emph{active} joint measurements.
However, while the latter probabilities are measured either by
actively inserting or removing a piece of nucleonic matter in the
beams, the measurement method via Eq.~(\ref{obs1}) is
completely different: a quantum--mechanical process alone
(namely, the dynamics of kaon decays) decides
if each one of the two kaons of a given pair is going to be
measured either in the strangeness or in the lifetime basis. The experimenter
remains totally passive in such measurements.

\subsection{Combining \textit{active} and \textit{passive} measurements}

One can similarly combine an \emph{active} measurement on one side with a \emph{passive}
measurements on the other. Let us consider, for instance, an
\emph{active} strangeness measurement on the left with outcome $K^0$
and a \emph{passive} lifetime or strangeness measurement on the right with outcome
$K_{f_r}=K_S$, $K_L$, $K^0$ or $\bar{K}^0$. One then has
to determine the relevant joint probabilities via the relation:
\begin{equation}
\label{obs2}
P\left[K^0(\tau_l),K_{f_r}(\tau_r)\right]=\frac{\Gamma[K^0, \tau_l ; f_r,\tau_r]}
{\Gamma(K_{f_r}\to f_r)\, N(\tau_l, \tau_r)} ,
\end{equation}
where:
\begin{equation}
\Gamma[K^0,\tau_l ; f_r,\tau_r]\equiv \int d\Omega_{f_r}\,
p[K^0,\tau_l;f_r,\tau_r] ,
\end{equation}
\begin{equation}
p[K^0, \tau_l ;f_r,\tau_r]\equiv \frac{1}{4} \biggl|e^{-i(\lambda_L \tau_l+\lambda_S \tau_r)}
\langle f_r|T|K_S\rangle_r -e^{-i(\lambda_S \tau_l+\lambda_L \tau_r)}
\langle f_r|T|K_L\rangle_r\biggl|^2 , \nonumber
\end{equation}
and the normalization factor, $N(\tau_l, \tau_r) \equiv
e^{-(\Gamma_L +\Gamma_S)(\tau_l +\tau_r)/2}
\cosh\left[\Delta\Gamma(\tau_l -\tau_r)/2\right]$, has the same
expression of the previous case with \emph{passive} joint
measurements. The normalization of the decay rate $\Gamma[K^0,
\tau_l ; f_r,\tau_r]$ reads $\sum_{f_r}\int d\tau_r \, \Gamma[K^0,
\tau_l ; f_r,\tau_r] = N(\tau_l,0)/2$.

As expected, the quantum--mechanical results of
Eqs.~(\ref{lSprob})--(\ref{probL}) are again reproduced.

\section{Quantum eraser experiments for kaons}\label{experiment}

Several quantum eraser experiments have already been performed
\cite{zeilinger95,scully00,Ts00,walborn,Tr02,Kim03}. They used
SPDC to produce a two--photon maximally entangled state which is
the analog of the kaon state of Eq.~(\ref{time}). One photon of
the pair is considered as the \emph{object} system. On this photon
one may obtain ``which way'' ($WW$) information by a suitable
measurement on the other photon, the \emph{meter}. Different
strategies ---quite often by exploiting photon
polarizations, as in Refs.~\cite{zeilinger95,walborn,Tr02,Kim03}---
are used for marking and subsequently erasing the meter $WW$
information. But
all these photon experiments need a kind of
double--slit mechanism in order to allow for a ``wave--like''
behaviour of the object system.
Moreover, they also require a meter marking procedure.

\begin{figure}
\begin{center}
\includegraphics[width=300pt, keepaspectratio=true]{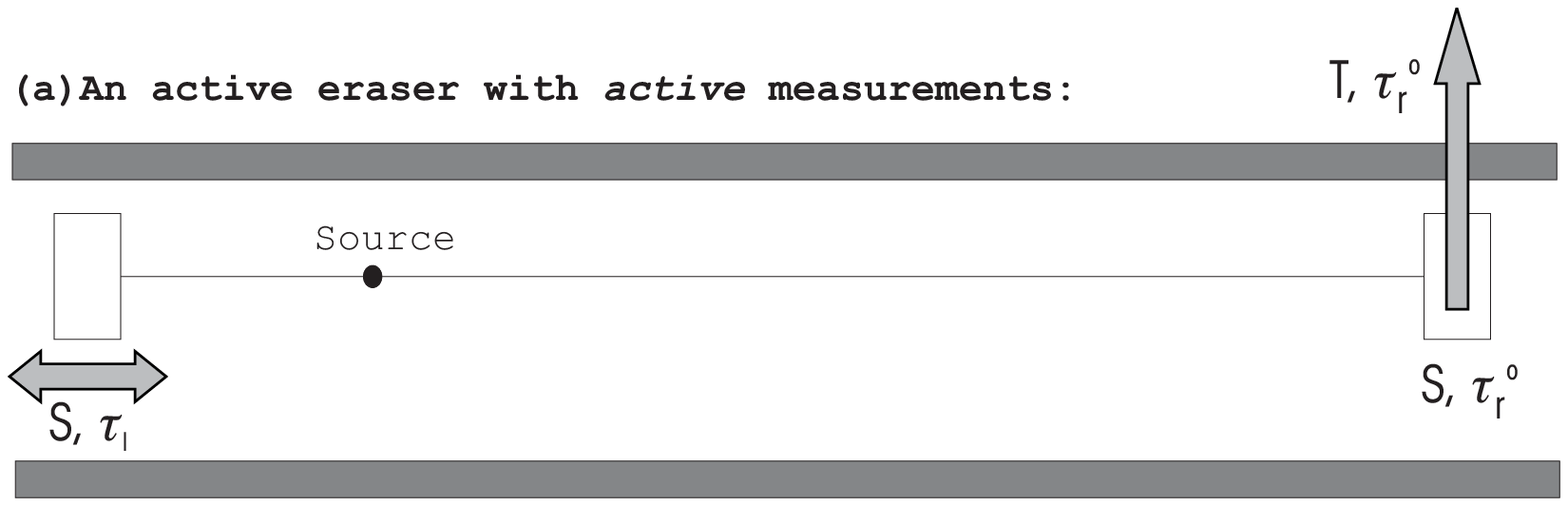}
\includegraphics[width=100pt, keepaspectratio=true]{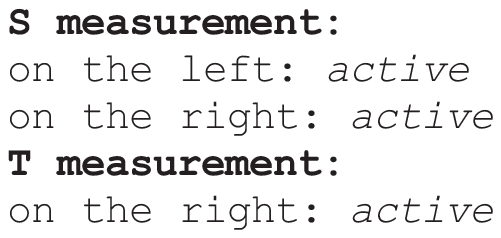}\\
\vspace{0.5cm}
\includegraphics[width=300pt, keepaspectratio=true]{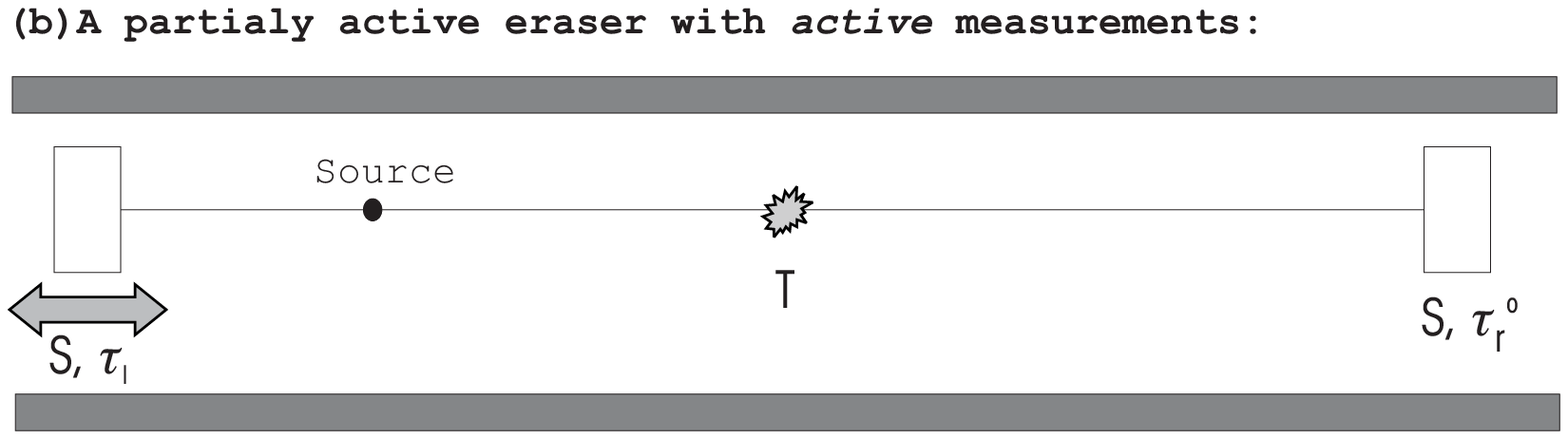}
\includegraphics[width=100pt, keepaspectratio=true]{QEa3.eps}\\
\vspace{0.5cm}
\includegraphics[width=300pt, keepaspectratio=true]{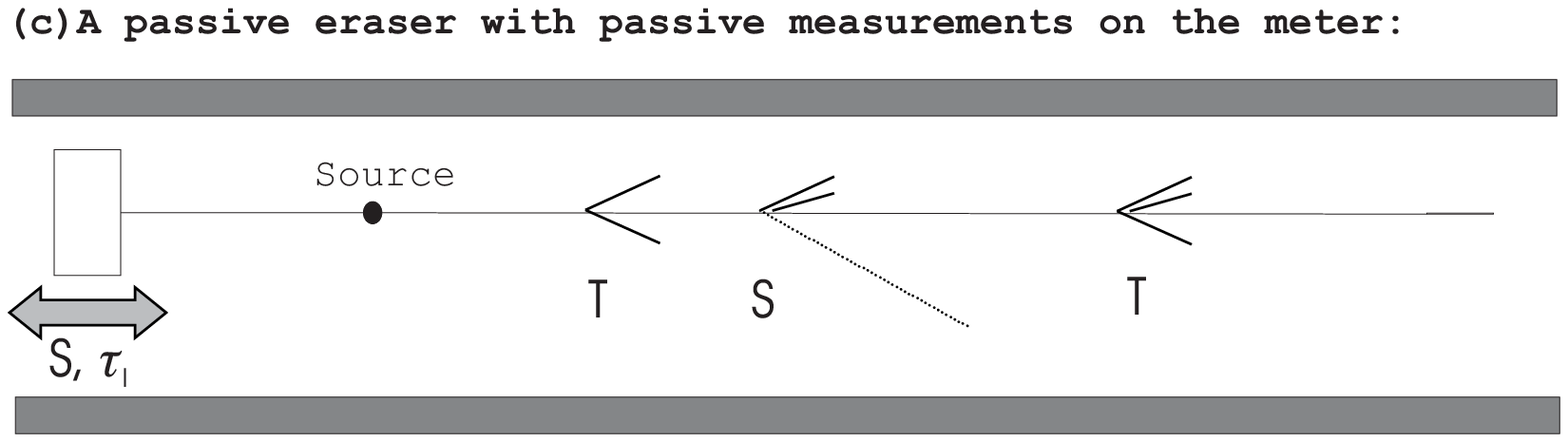}
\includegraphics[width=100pt, keepaspectratio=true]{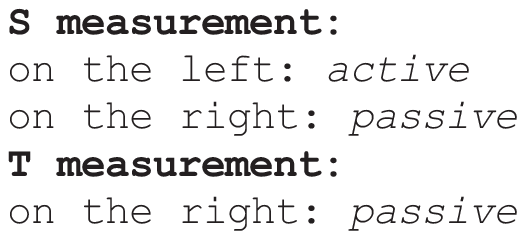}\\
\vspace{0.5cm}
\includegraphics[width=300pt, keepaspectratio=true]{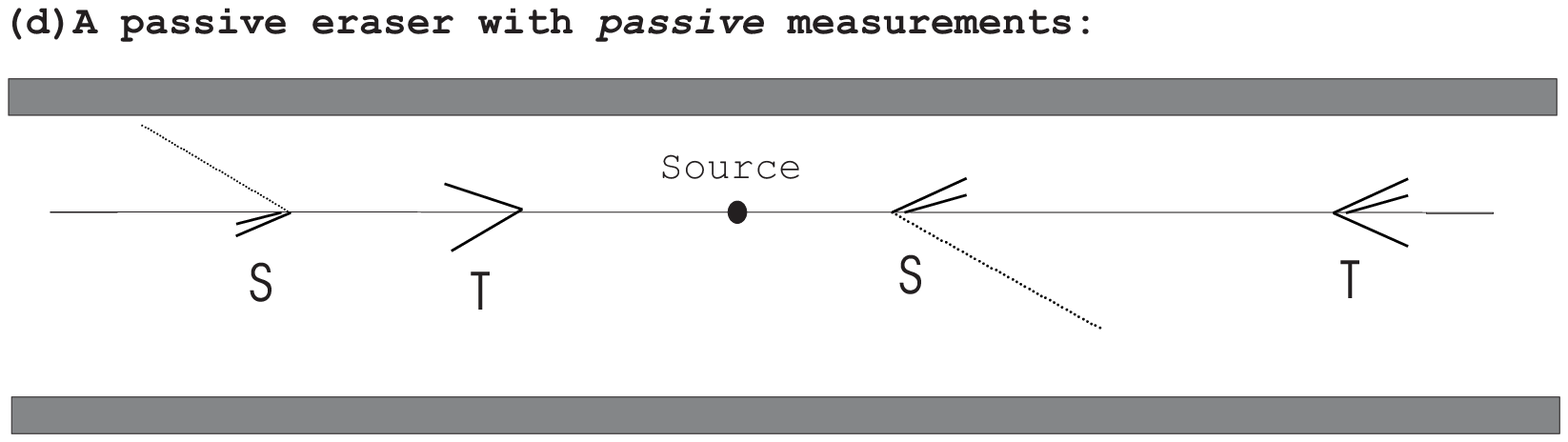}
\includegraphics[width=100pt, keepaspectratio=true]{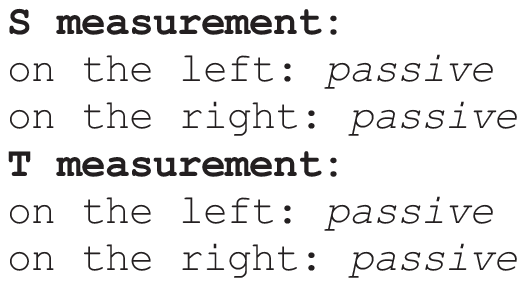}
\vspace{1cm}
  \caption{The figure shows four different experiments illustrating the
quantum marking and erasure procedures. In the first three experiments,
(a),(b) and (c), the object system propagates along the left, where
strangeness is \emph{actively} measured at various times $\tau_l$.
The meter system is on the right hand side. For the
last experiment, (d), it is not clear which side plays the role of the meter
since it is totally symmetric and it involves only
\emph{passive} measurements.}
\label{figure}
\end{center}
\end{figure}

By contrast, the two--kaon entangled state of Eq.~(\ref{time})
is given by Nature evolving through two amplitudes which are
automatically marked. One can then play the game of the quantum marking  and
erasure  experiments. Four possible experiments are illustrated in Fig.\ref{figure}
and discussed in the following. In
the first three cases, (a),(b) and (c), the left moving kaon is the
object system; on this kaon one performs \emph{active} strangeness
measurements at different $\tau_l$--values to
scan for possible strangeness oscillations. The right moving kaon is the
meter system; it carries ``which width'' ($W\mathcal{W}$) information which can
be actively (passively) erased by a suitable \emph{active} (\emph{passive})
measurement at a fixed time $\tau_r^0$ (at times below $\tau_r^0$).
In the fourth experiment, (d), \emph{passive} measurements are performed on both sides.
In this case, it is not clear which kaon plays the role of the meter.

\subsection*{(a) Active eraser with \textit{active} measurements}

In a first setup, strangeness detectors are \emph{actively}
inserted along each beam. Only kaon pairs which
survive up to both detectors are considered in this case.
We clearly observe $K_S$--$K_L$ interference
in the coincident counts of the object--meter system with a visibility
${\cal V}(\tau_l-\tau_r^0)$ given by Eq.~(\ref{visibility}).
More precisely, one expects fringes
for unlike--strangeness joint detections
[Eq.~(\ref{uSprob})] and
anti--fringes for like--strangeness joint detections
[Eq.~(\ref{lSprob})].
In a second setup, the strangeness detector previously placed along the right beam
is removed and only right going kaons that survive up to $\tau_r^0$ are considered. 
One then observes the lifetime of the meter and thus
obtains $W\mathcal{W}$ information
for the object kaon as well. The object--meter coincidence counts
do not exhibit interference: they follow the
non--oscillatory behaviour of Eqs.~(\ref{probS}) and (\ref{probL})
with $\Delta \tau = \tau_l-\tau_r^0$.

Hence, we have an experiment consisting of two setups in which the
experimenter can erase or not the $W\mathcal{W}$ information by
placing or not, at will, the strangeness detector along the right beam. The
first setup reveals the ``wave--like'' behaviour of the object kaon: if
$\tau_l = \tau_r^0$, the two propagating components $K_S$ and $K_L$
of the object are completely
indistinguishable because their marks are made totally inoperative by the
strangeness measurement on the meter kaon.
One gets interferences with maximal visibility as in conventional double--slit
experiments with indistinguishable paths.
However, for entangled neutral kaons the visibility
${\cal V}(\tau_l-\tau_r^0)$ of Eq.~(\ref{visibility}) decreases with
$|\tau_l-\tau_r^0|$ and vanishes when
$|\tau_l-\tau_r^0|\to \infty$; correspondingly, almost full $W\mathcal{W}$ information
is obtained for both kaons if $|\tau_l-\tau_r^0|\to \infty$. For a discussion
on a quantitative formulation of quantum erasure and the complementarity principle
for neutral kaons, see Refs.~\cite{BGH1,BGH2}.
The second setup  clearly demonstrates the ``particle--like'' behaviour of the object
kaon: no interference is observed because the meter mark is operative and one could
gain full $W\mathcal{W}$ information on the right moving kaon. This situation mimics
double--slit setups with complete $WW$ information.

This kaon experiment is analogous to the photon experiments
of Refs.~\cite{zeilinger95,Ts00,walborn}, as discussed in detail in Ref.~\cite{BGH1}.
Note, however, that  in the kaon case the amplitudes are automatically marked and
no double--slit device is needed.

\subsection*{(b) Partially active eraser with \textit{active} measurements}

In this case, a strangeness detector is always placed along the right beam
at a fixed time $\tau_r^0$. However, the experiment also detects and
considers eventual
decays of the right propagating kaon occurring between the origin and $\tau_r^0$.
In this way, it is the right moving kaon ---the meter--- that
eventually makes
the ``choice'' to show $W\mathcal{W}$ information by decaying before
$\tau_r^0$. If the meter kaon does indeed decay in free space, one measures its
lifetime \emph{actively}, i.e., via a decay time analysis,
and obtains $W\mathcal{W}$ information. If no decay is seen,
the incoming kaon is projected into one of the two strangeness states at
time $\tau_r^0$ by an \emph{active} strangeness measurement.

With this single experimental setup one observes the object
``wave--like'' behaviour (i.e., interference) for some events and
the ``particle--like'' behaviour (i.e., $W\mathcal{W}$
information) for others. The choice to obtain or not
$W\mathcal{W}$ information is naturally given by the instability
of the meter kaon. However, the experimenter can still choose when
---the time $\tau_r^0$--- he or she wants to detect the
strangeness of the surviving meter kaons, i.e., to erase the
object kaon $W\mathcal{W}$ information. Therefore, there is no
control over the marking and erasure operations for individual
kaon pairs, but probabilistic predictions for an ensemble of kaon
pairs are accessible.

This experiment is analogous to the eraser experiment with entangled
photons of Ref.~\cite{scully00}. The role played by the different beam--splitter
transmittivities in the photonic experiment is played by
the different $\tau_r^0$'s in the kaon case. If one chooses,
for instance, $\tau_r^0 \simeq 4.8 \; \tau_S$, one half of the events
correspond to right $K_{S}$ decays showing no oscillation and the
other half will show oscillations and antioscillations in joint
strangeness detections.

\subsection*{(c) Passive eraser with \textit{passive} measurement on the meter}

In this experiment one looks for the different decay modes in order
to identify the right moving meter kaon. This corresponds to a
\emph{passive} measurement of either strangeness or lifetime on
the meter. Now one clearly has a completely passive erasing
operation on the meter, and the experimenter has no control on the
operativeness of the lifetime mark. Only the object system is under
some kind of active control ---one still makes an \emph{active}
strangeness measurement on the object and considers only kaons
surviving up to this detector. Once the decay times of the right
kaons are taken into account, one recovers the same
joint probabilities as in the previous cases (a) and (b).

This experiment has no analog in any other considered
two--level quantum system.

\subsection*{(d) Passive eraser with \textit{passive} measurements}

In this experiment, both kaons evolve freely in space and the
experimenter observes, \emph{passively}, their decay modes. The
present procedure constitutes
the extreme case of a passive quantum eraser. The experimenter has no
control over individual pairs \emph{neither} on which of the two
complementary observables are measured \emph{nor} when
they are measured. The experiment is totally symmetric and thus it
shows the full behaviour of the maximally entangled state
(\ref{time}). However, since nothing is actively measured nor
erased, strictly speaking one cannot define this experiment as a standard quantum eraser.

Remarkably, the quantum--mechanical predictions for the observable
probabilities are again in agreement with those of all the previous experiments.
In particular, the joint probabilities for like-- and unlike--strangeness
passive measurements
coincide with those in Eqs.~(\ref{lSprob}) and (\ref{uSprob}). They are measured by
counting and properly sorting the two types of semileptonic decays
($\Delta S = \Delta Q = \pm 1$) at different values of $\tau_l$ and $\tau_r$.

Also this experiment has no analog in any other considered
two--level quantum system.

\section{Active and passive delayed choice}\label{delaychoice}

An essential and intriguing feature of the quantum eraser is that
it can be operated in the so called ``delayed choice'' mode
\cite{scully82,scully91}. If the meter is a system distinct and
spatially separated from the object, the decision to erase or not
the distinguishing mark on the meter can be taken even after the
potentially interfering object system has been detected. The
fact that everything works as in the ``normal'' time ordered mode
---in an apparently flagrant violation of causality---
has been a source of some controversy \cite{mohrhoff,scully99}.

Two specific features of our kaonic case help in this discussion:
(i) the two--kaon state is left--right (or, equivalently,
object--meter) symmetric, and (ii) time is not only the
characteristic parameter distinguishing the two operation modes
but also the variable in terms of which strangeness oscillations
are observed. The concurrence of these two circumstances in
quantum erasure for kaons allows for the following clarifying
discussion.

As previously described in the experiment (a) of the previous Section, 
on the right moving kaon one has the choice to
perform, \emph{actively}, either a strangeness or a lifetime measurement;
this is the meter system since it is measured at a fixed time
$\tau_r^0$. For the left moving kaon one measures strangeness
at a variable time $\tau_l$; this is then the
object system. If $\tau_{r}^0 \leq \tau_l$, one operates
in the normal, non--controversial mode: the left and right moving
kaons are the object and the meter systems, respectively, and the
meter mark is erased prior to the object strangeness measurement,
$\tau^0_{r\, (\rm meter)} \leq \tau_{l\, (\rm object)}$.

Let us consider now the ``delayed choice'' case in which at a given
$\tau_l$, such that $\tau_l <
\tau_r^0$, strangeness is measured \emph{actively} on the left
moving kaon.  Assume as our first case that one chooses to measure strangeness
on the right at $\tau_r^0$: at these given times $\tau_l$ and $\tau_r^0$
one can momentaneously invert the roles of the object and meter
kaons. One then operates as in the normal mode,
$\tau_{l\, (\rm meter)} < \tau^0_{r\, (\rm object)}$, and thus reobtains
the usual oscillations of Eqs.~(\ref{lSprob})--(\ref{visibility}) where both
$\cos(\Delta m\, \Delta \tau)$ and $\cosh(\Delta \Gamma \Delta\tau/2)$
are \emph{even} functions of $\Delta\tau$.
Since $\Delta\tau=\tau_l-\tau_r^0$ is a
time difference, one can then  allow for variations in $\tau_l$
and thus reconsider the left moving kaon as the object, oscillating
system. Assume in a second case that one chooses to measure \emph{actively}
lifetime instead of strangeness on the right at $\tau_r^0$: each
one of the four possible joint probabilities show the
non--oscillating feature. In both cases, the ``normal'' and ``delayed choice''
operation modes predict the same results.

It should be drawn to the reader's attention that the above
discussion applies to both \emph{active} and \emph{passive}
measurements on both the left and right moving kaons. The case in
which a \emph{passive} strangeness measurement is performed on the
meter provides a new concept of  ``delayed choice''  that we can
call a ``\emph{passive delayed choice}''.

More formally, for \emph{active} joint measurements we can discuss
again the issue in terms of projectors and time evolution
operators. The various joint probabilities can be computed as the
squared norms of the state
\begin{eqnarray}
\label{state1}
P^l P^r \underbrace{U_l(\tau_l,0)
U_r(\tau_r^0,0)|\phi(0)\rangle}_{|\phi(\tau_l,\tau_r^0)\rangle}\;,
\end{eqnarray}
where $P^{l,r}$ are the  four projectors corresponding to the
possible outcomes of each (left or right) measurement and
$U_{l,r}$ are the time evolution operators, which are unitary since we
retain and normalize to surviving kaon pairs. It is easy to show
\cite{GGW,BH1} that the state
\begin{eqnarray}
P^l U_l(\tau_l,\tau_r^0) P^r \underbrace{U_l(\tau_r^0,0)
U_r(\tau_r^0,0)|\phi(0)\rangle}_{|\phi(\tau_r^0,\tau_r^0)\rangle}
\end{eqnarray}
---with operators acting successively in the same order as in the
normal mode ($\tau_{r\, (\rm meter)}^0 \leq \tau_{l\, (\rm object)}$)---
has the same squared norm as the previous state (\ref{state1}).
This is so thanks to (i) the
commutativity of the operators acting on different Hilbert spaces
and (ii) the properties of unitary time evolution. In the case of
the ``delayed choice'' mode, the corresponding state
---with the appropriate ordering of the operators--- is
\begin{eqnarray}
P^r U_r(\tau_r^0, \tau_l) P^l \underbrace{U_l(\tau_l,0) U_r(\tau_l,0)
|\phi(0)\rangle}_{|\phi(\tau_l,\tau_l)\rangle}\;,
\end{eqnarray}
and has again the same squared norm as the previous two states.

\section{Conclusions}

Under the assumption of $CP$ conservation and the validity of the
$\Delta S = \Delta Q$ rule, we have shown that kaons admit two
alternative procedures to measure their two complementary
observables, strangeness and lifetime.  We call these procedures
\emph{active} and \emph{passive} measurements. The first one can
be seen as an analog to the usual von Neumann projection, while the
second one is clearly different and takes advantage of the
information spontaneously provided by neutral kaons through their decay
modes.

In this paper we propose four different experiments combining \emph{active} and
\emph{passive} measurement procedures in order to demonstrate the quantum erasure
principle for neutral kaons. All the considered experiments lead to the same
observable probabilities and ---more important for delayed
choice considerations--- this is true regardless of the temporal ordering of the
measurements. In our view, this illustrates the very nature of a quantum eraser
experiment: it essentially sorts different events, namely, strangeness--strangeness events
representing the ``wave--like'' property of the object
kaon or strangeness--lifetime events
representing the ``particle--like'' property of the object.
Time ordering considerations about the measurements are not relevant. The
same is true for the measurement methods: active measurements
---with the intervention of the experimenter--- and passive
measurements ---with no such intervention--- lead to equivalent
results.

Neutral kaon experiments verifying the
proposed quantum marking and eraser procedures has not been performed up to date.
As the only exception, the CPLEAR
collaboration \cite{CPLEAR} did part of the job required in our first
setup of experiment (a), showing the entanglement of kaon pairs from $p\bar p$
annihilation at rest through a measurement which
tested the oscillatory behaviour of strangeness--strangeness joint detections
for two values of $\tau_l-\tau_r$.
We think that the experiments proposed in this paper are of interest because
they offer a new test of the complementarity principle and shed new light on the very
concept of the quantum eraser.

\section*{Acknowledgements}

This work has been partly supported by EURIDICE
HPRN-CT-2002-00311, Spanish MCyT, BFM-2002-02588, 
Austrian Science Foundation (FWF)-SFB 015 P06 
and INFN. G.~G.~wishes to thank Prof.~R.~A.~Bertlmann and the Institute for 
Theoretical Physics, University of Vienna, for kind hospitality and financial
support.

\begin{thebibliography}{100}

\bibitem{scully82}
M. O.~Scully and K.~Dr$\ddot{\rm u}$hl, {\em Phys. Rev.} {\bf A 25}, 2208 (1982).


\bibitem{scully91}
M.~O.~Scully, B.-G.~Englert and H.~Walther, {\em Nature} {\bf 351}, 111 (1991).


\bibitem{Durr}
S.~D$\ddot{\rm u}$rr and G.~Rempe, {\em Opt. Commun.} {\bf 179}, 323 (2000).

\bibitem{zeilinger95}
T.~J.~Herzog, P.~G.~Kwiat, H.~Weinfurter and A.~Zeilinger,
{\em Phys. Rev. Lett.} {\bf 75}, 3034 (1995).


\bibitem{scully00}
Y.-H.~Kim, R.~Yu, S.~P.~Kulik, Y.~Shih and M.~O.~Scully,
{\em Phys. Rev. Lett.} {\bf 84}, 1 (2000).


\bibitem{Ts00}
T.~Tsegaye, G.~Bj$\ddot {\rm o}$rk, M.~Atat$\ddot {\rm u}$re, A.~V.~Sergienko,
B.~E.~A.~Saleh, and M.~C.~Teich, {\em Phys. Rev.} {\bf A 62}, 032106 (2000).


\bibitem{walborn}
S.~P.~Walborn, M.~O.~Terra Cunha, S.~P\'adua and C.~H.~Monken,
{\em Phys. Rev.} {\bf A 65}, 033818 (2002).


\bibitem{Tr02}
A.~Trifonov, G.~Bj$\ddot {\rm o}$rk, J.~S$\ddot {\rm o}$derholm and T.~Tsegaye,
{\em Eur. Phys. J.} {\bf D 18}, 251 (2002).


\bibitem{Kim03}
H.~Kim, J.~Ko, and T.~Kim, {\em Phys. Rev.} {\bf A 67}, 054102 (2003).


\bibitem{handbook}
For a survey on the physics at a ${\Phi}$--factory see
{\it The Second Da${\Phi}$ne Physics Handbook}, edited by L.~Maiani, G.~Pancheri
and N.~Paver (INFN, Laboratori Nazionali di Frascati, 1995).


\bibitem{CPLEAR}
A.~Apostolakis et al., {\em Phys. Lett.} {\bf B 422}, 339 (1998).


\bibitem{mohrhoff}
U.~Mohrhoff, {\em Am. J. Phys.} {\bf 64}, 1468 (1996); {\bf 67}, 330 (1999).


\bibitem{scully99}
M.~O.~Scully, B.-G.~Englert and H.~Walther, {\em Am. J. Phys.} {\bf 67}, 325 (1999).


\bibitem{bg1}
A.~Bramon and G.~Garbarino, {\em Phys. Rev. Lett.} {\bf 88}, 040403 (2002);
{\bf 89}, 160401 (2002);
A.~Bramon and M.~Nowakowski, {\em Phys. Rev. Lett.} {\bf 83}, 1 (1999).


\bibitem{ABN}
B.~Ancochea, A.~Bramon and M.~Nowakowski, {\em Phys. Rev.} {\bf D 60},
094008 (1999).


\bibitem{gen2004}
M.~Genovese, {\em Phys. Rev.} {\bf A 69}, 022103 (2004).


\bibitem{GGW}
G.~C.~Ghirardi, R.~Grassi and T.~Weber, Proceedings of the {\em Workshop on Physics
and Detectors for Da$\Phi$ne}, edited by G.~Pancheri (INFN, Laboratori Nazionali di
Frascati, 1991) p.261.

\bibitem{eberhard}
P.~H.~Eberhard, {\em Nucl. Phys.} {\bf B 398}, 155 (1993).


\bibitem{kaon-bell}
A.~Di Domenico, {\em Nucl. Phys.} {\bf B 450}, 293 (1995);
F.~Uchiyama, {\em Phys. Lett.} {\bf A 231}, 295 (1997);
F.~Benatti and R. Floreanini, {\em Phys. Rev.} {\bf D 57}, R1332 (1998);
R.~Foadi and F.~Selleri, {\em Phys. Rev.} {\bf A 61}, 012106 (2000);
N.~Gisin and A.~Go, {\em Am. J. Phys.} {\bf 69}, 264 (2001);
R.~Dalitz and G.~Garbarino,  {\em Nucl. Phys.} {\bf B 606}, 483 (2001);
M.~Genovese, C.~Novero and E.~Predazzi, {\em Phys. Lett.} {\bf B 513}, 401 (2001).
R.~A.~Bertlmann, W.~Grimus and B.~C.~Hiesmayr, {\em Phys. Lett.} {\bf A 289}, 21 (2001).

\bibitem{BH1}
R.~A.~Bertlmann and B.~C.~Hiesmayr, {\em Phys. Rev.} {\bf A 63},
062112 (2001); B.~C.~Hiesmayr, Ph.D. thesis, University of Vienna, 2002.


\bibitem{PDG}
Particle Data Group,
K.~Hagiwara et al., {\em Phys. Rev.} {\bf D 66}, 010001 (2002).


\bibitem{BGH1}
A.~Bramon, G.~Garbarino and B.~C.~Hiesmayr, {\em Phys. Rev. Lett.}
{\bf 92}, 020405 (2004).


\bibitem{kabir}
P.~K.~Kabir, {\it The CP Puzzle} (Academic Press, London, 1968).

\bibitem{BGH3}
A.~Bramon, G.~Garbarino and B.~C.~Hiesmayr, {\em Phys. Rev.}
{\bf A 69}, 022112 (2004).

\bibitem{CPLEARreview}
A.~Angelopoulos et al., {\em Phys. Rept.} {\bf 374}, 165 (2003);
{\em Phys. Lett.} {\bf 503}, 49 (2001): {\bf 444}, 38 (1998).


\bibitem{BGH2}
A.~Bramon, G.~Garbarino and B.~C.~Hiesmayr, {\em Eur. Phys. J.}
{\bf C 32}, 377  (2004).


\end{thebibliography}
\end{document}